\documentclass[runningheads]{llncs}

\usepackage[T1]{fontenc}
\usepackage{graphicx}
\usepackage{ifsym}
\usepackage[nocompress]{cite}
\usepackage[para,online,flushleft]{threeparttable}
\usepackage{amssymb}
\usepackage{multirow} 
\usepackage{booktabs} 
\usepackage{adjustbox}
\usepackage{array}
\usepackage[acronym]{glossaries}
\usepackage{multirow} 
\usepackage[ruled,linesnumbered,noend]{algorithm2e}
\usepackage{tikz}
\usepackage{siunitx}
\usepackage{pgfplots}
\usepackage{pgfplotstable}
\usepackage{csvsimple}
\usepackage{xcolor}
\usepackage{framed}
\usepackage{mdframed}

\usetikzlibrary{spy}
\pgfplotsset{compat=1.17}
\usepgfplotslibrary{%
	groupplots,
	external,
}
\tikzexternalenable

\newcommand\norm[1]{\lVert#1\rVert}

\newcommand{\drawcolorbar}{\tikzexternaldisable%
	\pgfplotscolorbardrawstandalone[
		scale=0.32, colormap={example}{samples of colormap = (8 of inferno)},
		colorbar horizontal,point meta max=0.2,colorbar style={ticks=none},
	]
\tikzexternalenable}

\begin{document}

\title{Product of Gaussian Mixture Diffusion Model for non-linear MRI Inversion}
\pgfplotsset{colormap={flare}{%
rgb = (0.92539502, 0.64345456, 0.47594352)
rgb = (0.92077582, 0.59804722, 0.44818634)
rgb = (0.9155979, 0.55210684, 0.42070204)
rgb = (0.90921368, 0.5056543, 0.39544411)
rgb = (0.90077904, 0.45884905, 0.37556121)
rgb = (0.888292, 0.40830288, 0.36223756)
rgb = (0.87199254, 0.3633634, 0.35974223)
rgb = (0.84916723, 0.32289973, 0.36711424)
rgb = (0.81942908, 0.28911553, 0.38102921)
rgb = (0.7826624, 0.26420493, 0.39754146)
rgb = (0.73695678, 0.24620072, 0.41357737)
rgb = (0.69226314, 0.23413578, 0.42480327)
rgb = (0.64795375, 0.22217149, 0.43330852)
rgb = (0.60407977, 0.21017746, 0.43913439)
rgb = (0.56041794, 0.19845221, 0.44207535)
rgb = (0.51278481, 0.18693492, 0.44112605)
rgb = (0.46818879, 0.17788392, 0.43552047)
rgb = (0.42355299, 0.16934709, 0.42581586)
rgb = (0.37928736, 0.16052483, 0.41270599)
rgb = (0.33604378, 0.15006017, 0.39835754)
}}

\pgfplotsset{colormap={inferno}{%
rgb = (1.46200e-03, 4.66000e-04, 1.38660e-02)
rgb = (2.94320e-02, 2.15030e-02, 1.14621e-01)
rgb = (9.29900e-02, 4.55830e-02, 2.34358e-01)
rgb = (1.83429e-01, 4.03290e-02, 3.54971e-01)
rgb = (2.71347e-01, 4.09220e-02, 4.11976e-01)
rgb = (3.60284e-01, 6.92470e-02, 4.31497e-01)
rgb = (4.41207e-01, 9.93380e-02, 4.31594e-01)
rgb = (5.28444e-01, 1.30341e-01, 4.18142e-01)
rgb = (6.09330e-01, 1.59474e-01, 3.93589e-01)
rgb = (6.94627e-01, 1.95021e-01, 3.54388e-01)
rgb = (7.69556e-01, 2.36077e-01, 3.07485e-01)
rgb = (8.41969e-01, 2.92933e-01, 2.48564e-01)
rgb = (8.98192e-01, 3.58911e-01, 1.88860e-01)
rgb = (9.44285e-01, 4.42772e-01, 1.20354e-01)
rgb = (9.72590e-01, 5.29798e-01, 5.33240e-02)
rgb = (9.86964e-01, 6.30485e-01, 3.09080e-02)
rgb = (9.84865e-01, 7.28427e-01, 1.20785e-01)
rgb = (9.66243e-01, 8.36191e-01, 2.61534e-01)
rgb = (9.46392e-01, 9.30761e-01, 4.42367e-01)
rgb = (9.88362e-01, 9.98364e-01, 6.44924e-01)
}}

\pgfplotsset{%
	marginlabels/.style={%
		ticklabel style={font=\fontsize{7pt}{8pt}\selectfont},
	},
	marginplot/.style={%
		height=.9\marginparwidth,
		width=.9\marginparwidth,
		scale only axis,
		axis y line=center,
		axis x line=middle,
		marginlabels,
	},
	cycle list={[indices of colormap={0,4,8,12,17} of flare]},
	pogmdm group plot/.style={%
		group/x descriptions at=edge bottom,
		group/y descriptions at=edge left,
		group/vertical sep=2mm,
		group/horizontal sep=2mm,
		width=1cm,
		height=1cm,
		scale only axis,
		no markers,
		ticklabel style={font=\tiny},
		grid=major,
		cycle list={[indices of colormap={0,4,8,12,17} of flare]},
		thin,
	},
}
\def\wwidth{2.7cm}
\def\hpad{2mm}
\def\vpad{0mm}
\pgfmathsetlengthmacro\spysize{\wwidth / 2 - 4 * 0.2pt}
\tikzset{
	mrispy/.style={%
		spy using outlines={%
			rectangle,
			magnification=3,
			width=\spysize,
			height=\spysize,
            color=orange
		}
	},
	font=\footnotesize,
}


\newacronym{mri}{MRI}{magnetic resonance imaging}
\newacronym{cs}{CS}{compressed sensing}
\newacronym{pi}{PI}{parallel imaging}
\newacronym{acl}{ACL}{auto calibration line}
\newacronym{cnn}{CNN}{convolutional neural network}
\newacronym{pogmdm}{PoGMDM}{product of Gaussian mixture diffusion model}
\newacronym{gan}{GAN}{generative adversarial network}
\newacronym{ebm}{EBM}{energy based model}
\newacronym{pde}{PDE}{partial differential equation}
\newacronym{corpd}{CORPD}{coronal proton density}
\newacronym{corpdfs}{CORPDFS}{coronal proton density fat suppressed}
\newacronym{rss}{RSS}{root sum of squares}
\newacronym{psnr}{PSNR}{peak signal-to-noise ratio}
\newacronym{ssim}{SSIM}{structural similarity}
\newacronym{nmse}{NMSE}{normalized mean squared error}
\newacronym{mmse}{MMSE}{minimum mean squared error}
\newacronym{map}{MAP}{maximum a posteriori}
\newacronym{ema}{EMA}{exponential moving average}
\newacronym{gmm}{GMM}{Gaussian mixture model}
\newacronym{foe}{FOE}{fields of expert}
\newacronym{tv}{TV}{total variation}
\newacronym{vn}{VN}{variational network}
\newacronym{ood}{OOD}{out of distribution}
\newacronym{id}{ID}{in distribution}
\newacronym{pc}{PC}{predictor corrector}
\newacronym{sde}{SDE}{stochastic differential equation}

\author{Laurenz Nagler\inst{1} \and
Martin Zach\inst{2}\orcidID{0000-0003-1941-875X} \and
Thomas Pock\inst{1}\orcidID{0000-0001-6120-1058}}

\authorrunning{L. Nagler et al.}

\institute{Technical University of Graz, 8010 Graz, Austria \\
\email{lnagler@student.tugraz.at}, \email{thomas.pock@icg.tugraz.at} \and
École polytechnique fédérale de Lausanne, 1015 Lausanne, Switzerland \\
\email{martin.zach@epfl.ch}}

\maketitle
\begin{abstract}
Diffusion models have recently shown remarkable results in magnetic resonance imaging reconstruction.
However, the employed networks typically are black-box estimators of the (smoothed) prior score with tens of millions of parameters, restricting interpretability and increasing reconstruction time.
Furthermore, parallel imaging reconstruction algorithms either rely on off-line coil sensitivity estimation, which is prone to misalignment and restricting sampling trajectories, or perform per-coil reconstruction, making the computational cost proportional to the number of coils.
To overcome this, we jointly reconstruct the image and the coil sensitivities using the lightweight, parameter-efficient, and interpretable product of Gaussian mixture diffusion model as an image prior and a classical smoothness priors on the coil sensitivities.
The proposed method delivers promising results while allowing for fast inference and demonstrating robustness to contrast out-of-distribution data and sampling trajectories, comparable to classical variational penalties such as total variation.
Finally, the probabilistic formulation allows the calculation of the posterior expectation and pixel-wise variance.
\end{abstract}
\section{Introduction}
\label{sec:introduction}
\Gls{mri} is one of the pillars of modern medicine, exhibiting high spatial resolution and contrast flexibility.
However, these strengths are limited by low temporal resolution, long scan times and artefacts induced by patient movement.

This can be overcome by acquiring less data, resulting in an ill-posed reconstruction problem.
One way to obtain solutions to the problem is \gls{pi}, where multiple spatially structured receiver coils are used.
The individual coil data is then combined in either the image domain~\cite{pruessmann1999SENSE} or the frequency domain~\cite{griswold2002GRAPPA}.
Another way of tackling this problem is through variational methods or \gls{cs} theory, which leverages sparse representations of the data in a specific domain and incoherent sampling to enable image reconstruction~\cite{block2007CSTV,lustig2007CSMRI}.

In recent years, data-driven methods have achieved remarkable results in \gls{mri} image reconstruction~\cite{chung2022scoreMRI,jalal2021robustCS,zach2023stabledeepmri}.
These models can be categorized by whether they follow a Bayesian separation of likelihood and prior, i.e.\ whether the learned components account for the measurement setup or not.
In the context of \gls{mri}, the Bayesian separation obviates the need for multi-coil data for learning~\cite{zach2023stabledeepmri} and enables the learned model to be used for reconstructing images from arbitrary k-space sampling trajectories.
Methods pursuing this Bayesian separation in \gls{mri} include works based on \glspl{gan}~\cite{narnhofer2019invGANS}, \glspl{ebm}~\cite{zach2023stabledeepmri}, and, more recently, diffusion models~\cite{jalal2021robustCS,song2022solvingINVProb,chung2022scoreMRI,uecker2022mri_markov}.
Diffusion models deliver state-of-the-art results but have shortcomings:
The networks are, in general, not the gradient of a potential~\cite{salimans2021shouldebmmodelscore}, and the time conditioning is opaque, seemingly unrelated to the underlying stochastic differential equation.
Furthermore, models with millions of parameters are often accompanied by long inference times, a lack of interpretability and the need for heuristics to be applied to data of arbitrary size~\cite{chung2022scoreMRI,erlacher2023jointnonlinearmriinversion}.
Existing methods for \gls{pi} such as~\cite{jalal2021robustCS,uecker2022mri_markov} typically require off-line coil sensitivity estimation, e.g. with ESPIRiT~\cite{uecker2014espirit}.
This is sensitive to patient motion and hard to adapt to arbitrary k-space sampling trajectories~\cite{knoll2011tgv_mri}.
In a different line of work,~\cite{chung2022scoreMRI} overcomes this by reconstructing individual coil images, making the computational cost proportional to the number of coils, which is unacceptably large when paired with the typical large networks used to model the score.
In~\cite{chung2023parallel}, the authors propose to tackle blind inverse problems by additionally learning a diffusion model on the parameters of the forward operator.
In the context of \gls{pi}, this translates to learning a diffusion model on the coil sensitivities, again making the computational cost proportional to the number of coils and introducing the engineering challenge of training and tuning the additional diffusion model.
In \cite{hu2024adobiadaptivediffusionbridge}, the authors propose an algorithm for \gls{pi}, where the coil sensitivities are assumed to be sufficiently close to an initial guess, again requiring off-line estimation.

Recently, Zach et al.~\cite{zach2024POGMDM} introduced the \gls{pogmdm},
\begin{equation}
\label{eq:model_eq}
    p_{\theta}(x, t) \propto \prod_{i,j=1}^{n,m} \prod_{k=1}^o \psi_k((K_k x)_{i,j}, w_k, t),
\end{equation}
a diffusion model with fields-of-experts-type structure and relations to classical shrinkage-type regularization.
Here, to model the density of \( n \times m \) images, for all \( k = 1, \dotsc, o \),
\begin{equation}
    \psi_k(x, w, t) = \sum_{i=1}^L w_i/\sqrt{2\pi\sigma(t)^2} \exp\bigl( (x - \mu_i)^2 / (2\sigma_k^2(t)) \bigr)
\end{equation}
is a \gls{gmm} with weights $w_k$, $K_k: \mathbb{R}^{n \times m} \rightarrow \mathbb{R}^{n \times m}$ is a convolution operator, and the time conditioning adapts the variance \( \sigma_k^2 \) in the \gls{gmm} in accordance with the \gls{sde} based on characteristics of \( K_k \).
This model is analytically tractable and interpretable, has few learnable parameters and allows for fast inference.
In their work, they only considered denoising due to the close relations between generative modelling and denoising via Tweedies identity.
In this work, we demonstrate the feasibility of \gls{pogmdm} as an image prior in joint non-linear \gls{mri} reconstruction by pairing it with simple smoothness priors on the coil sensitivities.

\section{Background}
\label{sec:background}
\subsection{Generative Modeling and Diffusion}
\label{ssec:gen_mod_diff}
In generative modeling, we are interested in learning a parametric model approximating a reference distribution from a given dataset.
Let $X$ be the random variable of the reference distribution with associated density $p_X$.
Learning a model $p_{\theta}$ approximating $p_X$ is hard as the high dimensional density space is sparsely populated.
One way to ease learning is to smooth (thus filling low-density regions) and approximate the density at different scales~\cite{song2019genmodelling}.

Song et al. generalized this idea to \glspl{sde}~\cite{song2021scorebased}.
In this work, we consider the \gls{sde}
\begin{equation}
\label{eq:diffusion_process}
    dY_t = \sqrt{2}dw_t, \quad Y_0 = X
\end{equation}
where $w$ is the standard Wiener process and is a random variable $Y_t$ with density $p_Y(\cdot, t)$.

The Fokker-Planck equation~\cite{Risken1996} links $Y_t$ to its density, which is given as the \gls{pde} $\partial p_Y(\cdot, t)/\partial t = \Delta p_Y(\cdot, t)$ with $p_Y(\cdot, 0) = p_X$.
This is the classic heat diffusion with the solution $p_Y(\cdot, t) = \mathcal{N}(0,2tI) * p_X$, thus constructing a scale space in the space of probability~\cite{zach2024POGMDM}.
In practice, a model $p_{\theta}(\cdot, t) \approx p_Y(\cdot, t)$ can be learned by optimizing the objective function
\begin{equation}
\label{eq:dns_objective}
    \min_{\theta} \int_0^{\infty} \mathbb{E}_{(x,y_t) \sim p_{X,Y_t}}\left[\norm{x - y_t - 2t\nabla p_{\theta}(y_t,t)}_2^2\right] dt
\end{equation}
known as denoising score matching~\cite{song2021scorebased,vincent2011dsm}.
With this configuration, sampling amounts to running the diffusion process,
\begin{equation}
\label{eq:revSDE}
    d\Bar{Y}_t = -2\nabla\log p_{Y}(\bar{Y_t}, t)dt +  \sqrt{2}d\bar{w}_t, \quad \Bar{Y}_T = Y_T,
\end{equation}
where $dt$ is a negative infinitesimal time step and $\bar{w}$ is the reverse time Wiener process starting, from a time $T$ to zero.
The only unknown quantity in \eqref{eq:revSDE} is the gradient of the log density $p_Y$ (score) at each time $t$, that we model with $p_{\theta}(\cdot, t)$.

\subsection{Inverse Problems and Diffusion Priors}
Solving an inverse problem amounts to recovering an unknown signal $x$ from a set of measurements  $z = \mathcal{A}(x) + \epsilon$, where $\mathcal{A}$ models the acquisition and \( \epsilon \) summarizes measurement noise.
The probabilistic treatment of the recovery problem amounts to constructing a posterior density $p_{X|Z}(\,\cdot\,,z)$ from the likelihood \( p_{Z|X} \), which makes use of the known acquisition physics encoded in \( \mathcal{A} \), and prior information about the reconstruction, encoded in \( p_X \); see~\cite{Dashti2017} for more details.
The diffusion framework can account for this by conditioning the stochastic process \eqref{eq:revSDE} on the measurements, resulting in
\begin{equation}
    d\Bar{Y}_t = -2\nabla\log p_{Y\mid Z}(\bar{Y_t}, z, t)dt +  \sqrt{2}d\bar{w}_t, \quad \Bar{Y}_T = Y_T,
    \label{eq:posterior reverse diffusion}
\end{equation}
where $\log p_{Y|Z}(y_t, z, t) \propto \log p_{Z \mid Y}(y_t, z, t) + \log p_{Y}(y_t, t)$.
The relationship between \( Y_t \) and \( Z \) is usually only known at \( t = 0 \), necessitating approximations for $t > 0$~\cite{jalal2021robustCS,song2022solvingINVProb,chung2023diffusion}.
\section{Methods}
\label{sec:methods}
The conditional reverse \gls{sde} \eqref{eq:posterior reverse diffusion} is valid for general, linear and nonlinear, inverse problems.
In this section, we state our acquisition model for \gls{pi} \gls{mri} and formulate our approximations.
The resulting reconstruction algorithm is summarized in Algorithm~\ref{alg:joint_recon}.
\subsection{Image Reconstruction Algorithm}
\label{ssec:img_recon_algo}
We view the \gls{pi} recovery problem as a non-linear inverse problem of jointly reconstructing the spin density $x \in \mathbb{C}^{n\times m}$ as well as the coil sensitivities $\sigma = (\sigma_1, \dots ,\sigma_{c}) \in \mathbb{C}^{n \times m \times c}$ from measured data $z = (z_1, \dotsc, z_{c}) \in \mathbb{C}^{f \times c}$ where $f \in \mathbb{N}$ is the number of measured spatial frequencies.
The relation between the data and the variables is given as 
\begin{equation}
\label{eq:non_lin_problem}
z = \mathcal{A}(x,\sigma) +\epsilon = 
    \begin{pmatrix}
        MF(\sigma_1 \odot x) \\
        \vdots\\
        MF(\sigma_{c} \odot x)
    \end{pmatrix}
    + \epsilon
\end{equation}
where $M: \mathbb{C}^{n \times m} \rightarrow \mathbb{C}^{f}$ is a binary sampling operator, $F: \mathbb{C}^{n \times m} \rightarrow \mathbb{C}^{n \times m}$ is the discrete Fourier transform and $\epsilon \in \mathbb{C}^{f\times c}$ is additive Gaussian noise.
Recovering $(x, \sigma)$ from \( z \) is hard even in the fully sampled case~\cite{uecker2008joinrecon}.
To approach this, we introduce priors on the spin density and the sensitivities:
Motivated by the bilinear form of~\eqref{eq:non_lin_problem}, we view the recovery problem as a blind inverse problem~\cite{chung2023parallel,hu2024adobiadaptivediffusionbridge} and utilize the approximation
\begin{equation}
\label{eq:full posterior}
p_{X,\Sigma|Z}(x_t, \sigma_t, z, t) \propto p_{Z|X,\Sigma}(x_t, \sigma_t, z, t)\cdot p_{\theta}(x_t, t)\cdot p_{\Sigma}(\sigma_t, t).
\end{equation}
i.e.\ that the distribution of \( (X, \Sigma) \) factorizes.
We choose a Gaussian likelihood of the form $p_{Z|X,\Sigma}(x_t, \sigma_t, z, t) \propto \exp\left(-\tfrac{1}{2}\norm{\mathcal{A}(x_t,\sigma_t) - z}_2^2\right)$.
This is a popular and simple choice originating from~\cite{jalal2021robustCS}, but only correct for $t=0$ in the diffusion process~\cite{chung2022scoreMRI}.
$p_{\theta}(\cdot, t)$ is the \gls{pogmdm} \eqref{eq:model_eq} trained on reference images, see the details in section \ref{ssec:details}.
Following~\cite{zach2023stabledeepmri} we chose, the classical smoothness prior $p_{\Sigma}(\sigma, 0) \propto \exp(-s(\sigma))$, with
\begin{equation}
\label{eq:smoothness_prior}
    s:\mathbb{C}^{n \times m \times c} \rightarrow \mathbb{R}:
    \sigma \mapsto \frac{1}{2}\sum_{i=1}^{c} \left( \norm{D_\text{D}\text{Re}(\sigma_{i})}_2^2 + \norm{D_\text{D}\text{Im}(\sigma_{i})}_2^2\right),
\end{equation}
where $D_\text{D}:\mathbb{R}^{n \times m} \rightarrow \mathbb{R}^{n \times m \times 2}$ is a forward finite differences operator with Dirichlet boundary conditions.
We use an adapted version of the algorithm proposed in~\cite{erlacher2023jointnonlinearmriinversion} for posterior sampling.
The likelihood is incorporated by doing a gradient descent step in each iteration, where the gradient of the log-likelihood w.r.t.~$x_t$ is 
\begin{equation}
\label{eq:grad_x_liklihood}
    \nabla_{x_t} \log p_{Z|X,\Sigma}(x_t, \sigma_{t}, y, t) = \sum_{i=1}^{c} \bar{\sigma}_{t,i} \odot (F^*M^*(MF(\sigma_{t,i} \odot x_t) - z_i)), 
\end{equation}
with $\bar{\sigma}_{t,i}$ being the complex conjugate sensitivity of the coil indexed by $i$.
We found it sufficient to fix \( p_\Sigma(\,\cdot\,,t) = p_\Sigma(\,\cdot\,,0) \) for all \( t > 0 \), thus depart from the diffusion framework and perform proximal gradient descent steps on $\sigma_t$.
The proximal map of $\log p_{\Sigma}$ has a closed-form solution 
\begin{equation}
\label{eq:sens_prox}
\text{prox}_{\mu s}(\sigma) = 
\begin{pmatrix}
    Q_\mu(\text{Re}(\sigma_{1})) + iQ_\mu(\text{Im}(\sigma_{1})) \\
    \vdots \\
    Q_\mu(\text{Re}(\sigma_{c})) + iQ_\mu(\text{Im}(\sigma_{c}))
\end{pmatrix},
\end{equation}
with $\mu \in \mathbb{R}_{>0}$ and $Q_\mu:\mathbb{R}^{n \times m} \rightarrow \mathbb{R}^{n \times m}; x \mapsto \mathcal{S}^*\left(\mathcal{S}(\mu x) \odot (\tau + \mu)^{-1}\right)$.
Here, $\mathcal{S}$ is the two-dimensional discrete sine transform, and $\tau$ are the eigenvalues of the two-dimensional discrete Laplace operator~\cite{zach2023stabledeepmri}.
The gradient of the log-likelihood w.r.t. $\sigma_t$ is 
\begin{equation}
\label{eq:grad_sigma_likelihood}
\nabla_{\sigma_t} \log p_{Z|X,\Sigma}(x_t, \sigma_t, y, t) = 
\begin{pmatrix}
    x_t \odot (F^*M^*(MF(\sigma_{t,1} \odot x_t) - z_1)) \\
    \vdots \\
    x_t \odot (F^*M^*(MF(\sigma_{t,c} \odot x_t) - z_{c} )) 
\end{pmatrix}.
\end{equation}

\begin{algorithm}
    \DontPrintSemicolon
    \SetKwInOut{Output}{Output}

    \Output{$x^0$, $\sigma^0$}

    \( x^N \sim \mathcal{N}(0, \zeta_{max}^2I) \)
    
    \For{\( i \in [N-1, 0] \)}{
        \For{\( k \in [\textnormal{Re}, \textnormal{Im}] \)}{
            \( x^{k,i} =  x^{k,i+1} + (\zeta_{i+1}^2 - \zeta_{i}^2)\nabla\log p_{\theta}(x^{k,i+1},\zeta^2_{i+1}) + \sqrt{\zeta_{i+1}^2 - \zeta_{i}^2}\xi_i^{k} \)\;
        }
        \( x^{i} = x^{\text{Re},i} + \text{i}x^{\text{Im},i}\)\;
        \( x^{i} = x^{i} + \lambda\nabla_{x^i}\log p_{Z|X,\Sigma}(x^i, \sigma^{i+1}, z, t)\)\;
        
        \For{\( j \in [0, M-1] \)}{
            \For{\( k \in [\textnormal{Re}, \textnormal{Im}] \)}{
                \( x^{k,i} =  x^{k,i} + \epsilon_i \nabla\log p_{\theta}(x^{k,i}, \zeta^2_{i}) + \sqrt{2\epsilon_i}\xi_j^{k}\)\;
            }
            \( x^{i} = x^{\text{Re},i} + \text{i}x^{\text{Im},i}\)\;
            \( x^{i} = x^{i} + \lambda\nabla_{x^i}\log p_{Z|X,\Sigma}(x^i, \sigma^{i+1}, z, t)\)\;
        }
        \( \sigma^i = \text{prox}_{\mu s}\bigl(\sigma^{i+1} - \mu\nabla_{\sigma^{i+1}} \log p_{Z|X,\Sigma}(x^i, \sigma^{i+1}, z)\bigr) \)
    }
    \caption{Joint Reconstruction Algorithm}
    \label{alg:joint_recon}
\end{algorithm}
Similar to~\cite{chung2022scoreMRI}, we apply the prior, trained on magnitude images, to the real- and imaginary parts individually.
We summarize the joint reconstruction in Algorithm \ref{alg:joint_recon}, where $\xi$ stores independent samples of a standard complex-normal distribution.

\subsection{Model Architecture and Implementational Details}
\label{ssec:details}
We largely use the same model parameterization as in~\cite{zach2024POGMDM}:
The convolution operators \( \{ K_k = \gamma_k \tilde{K}_k \}_{k=1}^o \) are nonseparable shearlets \( \tilde{K}_k \) comprising two scales with five shearings for each of the two (horizontal and vertical) cones, yielding $o=20$, each endowed with a learnable weight $\gamma_{k} \geq 0$.
Following~\cite{zach2024POGMDM,bogensperger2022piggyback}, we learn the construction blocks of the shearlet system, a one-dimensional low-pass-filter $h \in \mathbb{R}^9$ and a two-dimensional directional high-pass filter $P \in \mathbb{R}^{17\times 17}$.
For each \( k = 1, \dotsc, o \), we chose $\psi_k : \mathbb{R} \times \triangle^L \times \mathbb{R}_{\ge0} \rightarrow \mathbb{R}_{\ge0}$ as an $L=125$ component \gls{gmm} whose weights $w_k$ are constrained to the $L$-dimensional unit simplex $\triangle^L$.
This results in \( 9 + 17^2 + o\lceil L/2 \rceil + o = 1578 \) (as in~\cite{zach2024POGMDM} the number of learnable weights is half the number of components due to symmetry) learnable parameters that are optimized for \eqref{eq:dns_objective} with projected AdaBelief~\cite{zhuang2020adabelief} for 100 000 steps, where we implement the same constraints as in~\cite{zach2024POGMDM}.
Furthermore, we use \gls{ema}, with a momentum of 0.999.

For posterior sampling we choose $N=1000$, $M=1$, $\lambda=1$, $\mu=10$ and the noise schedule $\zeta(t) = \zeta_{\mathrm{min}}(\zeta_{\mathrm{max}}/\zeta_{\mathrm{min}})^t$ with $\zeta_{\mathrm{min}} =0.01$ and $\zeta_{\mathrm{max}} = 10$. We calculate $\epsilon_i$ according to~\cite{song2021scorebased}. 
Furthermore, we find optimal hyperparameters for each sampling pattern with grid search.
The coil sensitivities $\sigma$ are initialized with the zero-filled coil images normalized by the initial \gls{rss} reconstruction.
Finally, we accelerate sampling by using the algorithm proposed in~\cite{chung2022ccdf} with vanilla initialization.

\subsection{Experimental Data}
\label{ssec:exp_data}
We use the fastMRI knee dataset~\cite{knoll2020fastmri} for model training, hyper-parameter search, and evaluation.
We use the central eleven slices of each scan in the \gls{corpd} training split, resulting in 5324 images of size $320 \times 320$.
For the hyper-parameter search and testing, we divide the validation split into 30 validation files and 58 test files, excluding k-space data with width different from 368 and 372 for simplicity of implementation, again taking the central eleven slices.
For \gls{ood} experiments we use the \gls{corpdfs} validation dataset, again excluding based on the width and using the eleven central slices.
Training images are normalized to a maximum of 1 via $x \mapsto x / \norm{x}_{\infty}$.

\subsection{Comparison and Evaluation}
\label{ssec:comp_eval_methods}
Our evaluation is twofold: Firstly, we conduct a synthetic single coil experiment, where we retrospectively sample the k-space data of \gls{corpd} reference \gls{rss} images $x \in \mathbb{R}^{320 \times 320}$.
Secondly, a \gls{pi} experiments with joint reconstruction on the \gls{corpd}/FS datasets.
In both experiments, we compare against the Charbonnier smoothed isotropic \gls{tv} as a classical variational penalty in the joint nonlinear inversion algorithm proposed in~\cite{zach2023stabledeepmri}.
ScoreMRI~\cite{chung2022scoreMRI} (single coil), the fastMRI baseline~\cite{zbontar2019fastmriopendatasetbenchmarks} (single coil), and the end-to-end \gls{vn}~\cite{sriram2020EndToEndVN} (\gls{pi}) serve as state-of-the-art references, where the latter two primarily serve as examples of possible pitfalls in signal recovery without proper Bayesian separation.
We did not include ScoreMRI in the \gls{pi} experiments as it can only handle fixed-size data ($320 \times 320$).

We estimate the \gls{mmse} by averaging ten posterior samples and, similar to~\cite{uecker2022mri_markov}, find the \gls{map} by performing 250 accelerated gradient descent~\cite{nesterov1983agd} steps with a fixed $\sigma$ and a step size of 0.001 after the full reverse diffusion.
For quantitative comparison, we calculate the \gls{psnr}, \gls{ssim}, and \gls{nmse} with respect to the fully-samples RSS reconstruction.
To quantitatively compare the reconstruction to other \gls{rss} reconstructions, we weigh the reconstruction by the \gls{rss} of the coils as described in \cite{uecker2008joinrecon}, and we follow \cite{zach2023stabledeepmri} and fit a spline curve to match the reconstructed and reference intensities.
\section{Results}
\label{sec:results}
\label{ssec:Overcomplete Model}
\begin{figure*}
	\begin{tikzpicture}
		\begin{filecontents*}{figures/shearlets/lamdas.csv}
0.36
2.52
1.92
2.53
0.36
1.26
1.10
1.56
1.08
1.27
1.62
0.40
2.17
0.40
1.64
1.02
1.16
1.47
1.17
1.02
\end{filecontents*}
\csvreader[no head]{figures/shearlets/lamdas.csv}{1=\llambda}{
			\pgfmathsetmacro{\component}{int(\thecsvrow-1)}
			\pgfmathsetmacro{\yy}{-int(\component/10)*1.2}
			\pgfmathsetmacro{\xx}{mod(\component,10)*1.15}
			\node at (\xx-0.55, \yy-.6) {\tiny \( \llambda \)};
			\node at (\xx-0.55, \yy) {\includegraphics[width=1.00cm]{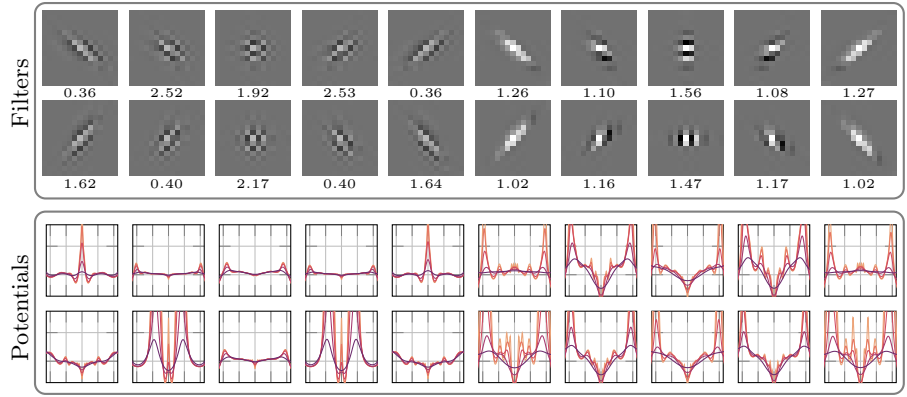}};
		}
		\draw [gray, thick, rounded corners] (-1.15cm, -2cm) rectangle (10.4cm, 0.6cm);
		\node at (-1.35cm, -0.6cm) [rotate=90] {Filters};

        \begin{scope}[xshift=-1cm,yshift=-3.3cm]
			\begin{groupplot}[
                pogmdm group plot,
				group style={
					group size=10 by 2,
				},
				width=0.95cm,
				height=0.95cm,
				ymin=-1.5,
				ymax=3.5,
				xmin=-0.45,
				xmax=0.45,
                xticklabel=\empty,
                yticklabel=\empty
			]
				\pgfplotsinvokeforeach{0,...,19}
				{%
					\nextgroupplot%
					\pgfplotsforeachungrouped \ccol in {a,b,c,d,e}
					{%
						\edef\tmp{%
							\noexpand\addplot table[col sep=comma, x=x, y=\ccol]{figures/shearlets/#1/potentials.csv};%
						}\tmp%
					}%
				}

			\end{groupplot}
			\draw [gray, thick, rounded corners] (-0.15cm, -1.3cm) rectangle (11.4cm, 1.125cm);
			\node at (-0.35cm, -0.15cm) [rotate=90] {Potentials};
		\end{scope}
	\end{tikzpicture}
    \caption{The learned overcomplete model with Gaussian mixture experts.
Below each filter the weight $\gamma_k$ is shown.
The first row shows the vertical cone and the second row is the horizontal cone of the shearlet system.
The first five entries correspond to the shearing for the first scale and the second five for the second scale.
The colours indicate the diffusion time 
    $
    \sqrt{\num{2}t} = 
    \num{0} \protect\tikz[baseline=-\the\dimexpr\fontdimen22\textfont2\relax]\protect\draw [index of colormap={0} of flare, thick] (0,0) -- (.5, 0);,
    \num{0.025} \protect\tikz[baseline=-\the\dimexpr\fontdimen22\textfont2\relax]\protect\draw [index of colormap={4} of flare, thick] (0,0) -- (.5, 0);,
    \num{0.05} \protect\tikz[baseline=-\the\dimexpr\fontdimen22\textfont2\relax]\protect\draw [index of colormap={8} of flare, thick] (0,0) -- (.5, 0);,
    \num{0.1} \protect\tikz[baseline=-\the\dimexpr\fontdimen22\textfont2\relax]\protect\draw [index of colormap={12} of flare, thick] (0,0) -- (.5, 0);,
    \num{0.2} \protect\tikz[baseline=-\the\dimexpr\fontdimen22\textfont2\relax]\protect\draw [index of colormap={17} of flare, thick] (0,0) -- (.5, 0);
    $.}
    \label{fig:learned_model}
\end{figure*}

We demonstrate the interpretability of the image prior by showing the learned filters $\tilde{K}_k$, weights $w_k$, and potentials $-\log\psi_k(\cdot, w_k, t)$ for different $t$ in Figure \ref{fig:learned_model}.
Due to the overcompleteness of the model, the potentials significantly differ from the leptokurtic filter responses observed in~\cite{huang1999imagestatistics}, instead showing multiple minima different from zero, allowing the enhancement of certain image structures.
This matches the observations from~\cite{zach2024POGMDM,pock2017tnrd,zhu1997priorlearning}. 

\subsection{Reconstruction}
\label{ssec:reconstruction}

Table \ref{tab:single_coil} shows the quantitative results for the synthetic single coil experiment.
As expected from the large, resource-intensive model, ScoreMRI~\cite{chung2022scoreMRI}  performs best across all sampling trajectories.
The performance of the UNet, which does not follow a strictly Bayesian separation, significantly degrades for sampling trajectories different from the training setup (Cartesian with an acceleration of four).
In contrast to other learning-based methods, our model has relatively few learnable parameters while achieving satisfactory results, outperforming \gls{tv} across all sampling trajectories, except for the radial pattern, where it performs equally well.

\begin{table}
    \centering
    \caption{
        Quantitative results for the single-coil experiment.
        Bold numbers represent the best results, and underlined numbers are the second-best.
        The \gls{nmse} is scaled by $10^2$.
    }
    \begin{center}
    \begin{threeparttable}
    \begin{tabular}{cccccccc}
        \toprule
        Pattern & A & Metric &  ZF & TV & UNet & ScoreMRI &  Ours (\gls{mmse})  \\
        \midrule
        \multirow{3}{*}{Cartesian}
        & \multirow{3}{*}{4} & \gls{psnr} & 24.03 & 30.2 & $\underline{33.85}$ & $\mathbf{34.74}$ & 31.66  \\
        & & \gls{ssim} & 0.69 & 0.81 & $\mathbf{0.87}$ & $\mathbf{0.87}$ & $\underline{0.83}$ \\
        & & \gls{nmse} & 6.32 & 1.23 & $\underline{0.48}$ & $\mathbf{0.39}$ & 0.84  \\
        \midrule
        \multirow{3}{*}{Spiral}
        & \multirow{3}{*}{$\approx 5$} &  & 21.15 & 31.32 & 27.81 & $\mathbf{35.67}$ & $\underline{32.99}$  \\
        & &  & 0.63 & 0.84 & 0.78 & $\mathbf{0.88}$ & $\underline{0.86}$\\
        & &  & 10.61 & 0.91 & 1.91 & $\mathbf{0.31}$ & $\underline{0.63}$ \\
        \midrule
        \multirow{3}{*}{Radial}
        & \multirow{3}{*}{$\approx 6$} &  & 24.6 & $\underline{31.34}$ & 29.85 & $\mathbf{33.62}$ & $\underline{31.34}$    \\
        & &  & 0.68 & $\underline{0.82}$ & 0.79 & $\mathbf{0.84}$ & $\underline{0.82}$ \\
        & &  & 5.33 & $\underline{0.89}$ & 1.19 & $\mathbf{0.50}$ & $\underline{0.89}$ \\
        \midrule
        \multirow{3}{*}{2D Gaussian}
        & \multirow{3}{*}{8} &  & 21.77 & 30.40 & 24.34 & $\mathbf{34.22}$ & $\underline{33.19}$  \\
        & &  & 0.65 & $\underline{0.85}$ & 0.73 & $\underline{0.85}$ & $\mathbf{0.86}$ \\
        & &  & 7.90 & 1.38 & 4.35 & $\mathbf{0.44}$ & $\underline{0.64}$ \\
        \midrule
        \multicolumn{3}{c}{Learnable parameters} & - & - & $4.9\times10^8$ & $6.7\times10^{7}$ & $1578$ \\
        \bottomrule
    \end{tabular}
    \begin{tablenotes}
        A: Acceleration
    \end{tablenotes}
    \end{threeparttable}
    \end{center}
    \label{tab:single_coil}
\end{table} 

\begin{figure*}[!ht]
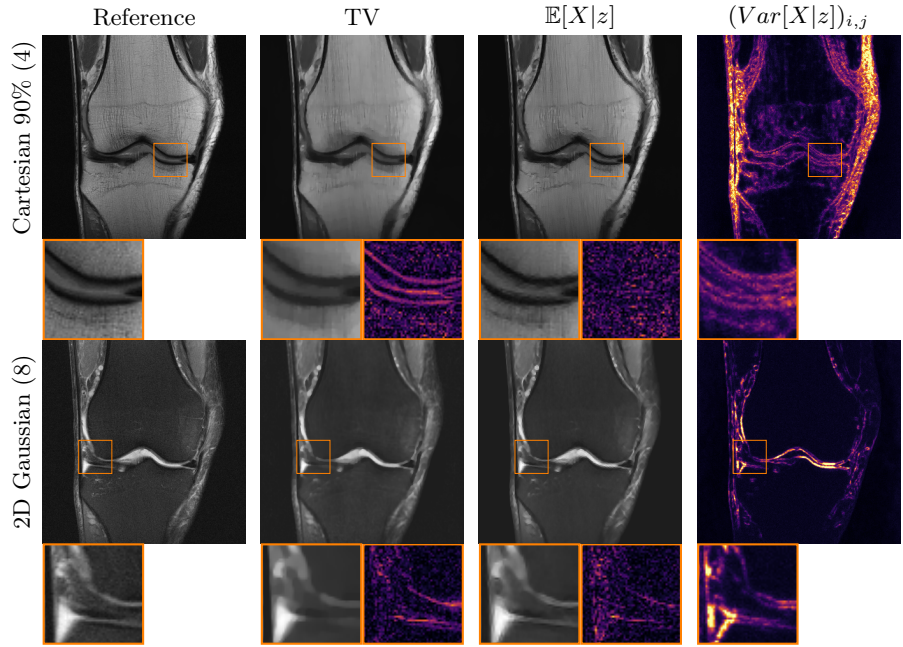

	\centering
	\tikzexternaldisable
	\begin{tikzpicture}
		\def\spyoff{{{0.35cm,-0.3cm},{-0.65cm,-0.2cm}}}
		\foreach [count=\isampling] \sampling/\sanno in {
			cartesian_rot90/Cartesian 90\% (4),
            gaussian2d/2D Gaussian (8)%
		}{
			\pgfmathsetlengthmacro{\yy}{-\isampling * (\wwidth + \wwidth / 2 + \vpad)}
			\pgfmathsetmacro{\spyxoff}{\spyoff[\isampling-1][0]}
			\pgfmathsetmacro{\spyyoff}{\spyoff[\isampling-1][1]}
			\foreach [count=\imethod] \method/\manno in {%
                reference/Reference,
				tv/TV,
				mmse/$\mathbb{E}[X|z]$,
				var_c/$(Var[X|z])_{i,j}$%
			}{
				\pgfmathsetlengthmacro{\xx}{\imethod * (\wwidth + \hpad)}
				\ifthenelse{\isampling=1}{\node at (\xx, -\wwidth / 2 - 1.1cm) {\manno};}{}
                \ifthenelse{\isampling=2}{\def\ds{corpdfs}}{\def\ds{corpd}}
				\pgfmathsetlengthmacro{\xxspy}{\xx + \spyxoff}
				\pgfmathsetlengthmacro{\yyspy}{\yy + \spyyoff}
				\coordinate (onn) at (\xxspy, \yyspy);
				\ifthenelse{\imethod=1 \OR \imethod=4}{}{
				\begin{scope}[mrispy]
					\node [inner sep=0, outer sep=0] at (\xx, \yy) {\includegraphics[angle=0,origin=c,width=\wwidth]{figures/\sampling/\ds/d_\method_c.png}};
					\spy on (onn) in node at (\xx + \wwidth / 2 / 2, \yy - 1.5 * \wwidth / 2);
				\end{scope}}
				\begin{scope}[mrispy]
					\node [inner sep=0, outer sep=0] at (\xx, \yy) {\includegraphics[angle=0,origin=c,width=\wwidth]{figures/\sampling/\ds/\method.png}};
					\spy on (onn) in node at (\xx - \wwidth / 2 / 2, \yy - 1.5 * \wwidth / 2);
				\end{scope}
			}
            \node [rotate=90, overlay] at (1.3, \yy) {\sanno};
		}
	\end{tikzpicture}
	\tikzexternalenable
	\label{fig:parallel_imaging}
    \caption{Qualitative comparison for \gls{pi} against \gls{tv}.
The first row shows results on \gls{id} data, the second on \gls{ood} data.
The zoom shows an image region and the corresponding absolute error and the pixel-wise variance in the right column (\num{0}~\protect\drawcolorbar~\num{0.25}).}
\end{figure*}

The quantitative results for the joint nonlinear inversion in Table \ref{tab:pi_corpd_fs} demonstrate that our method outperforms \gls{tv} on CORPD data on all sampling patterns except for Gaussian, where the performance is comparable.
On CORPDFS data, our method is second to \gls{tv} for most sampling trajectories, which we believe is due to the noise in the reference images, impairing quantitative comparison.
This is supported by the qualitative results in Figure \ref{fig:parallel_imaging}, where the error maps in the insets show that \gls{tv} introduces significant errors in the anatomy, e.g. removing the cartilage from the tibia in the CORPDFS reconstruction, or the blood vessels in the lateral region.
In contrast, our reconstruction is able to retain these features while delivering an artifact-free reconstruction.
As in the single coil case, the end-to-end \gls{vn} only performs well in its training configuration.

Without adding any extra computational cost to the MMSE estimate, the diffusion framework provides valuable insights into reconstruction uncertainty by delivering pixel-wise marginal variances, as shown in \ref{fig:parallel_imaging}.
Reconstructing an image of size $640 \times 372$ with $16$ coils takes twenty seconds with our algorithm.
In contrast, ScoreMRI~\cite{chung2022scoreMRI} takes approximately five minutes to reconstruct an image of size $320 \times 320$ from single-coil data on the same hardware.

\begin{table}
\centering
\caption{Quantitative results for \gls{pi} experiments with different sampling trajectories.
The rows in each pattern alternate between \gls{psnr}, \gls{ssim} and \gls{nmse}.
The \gls{nmse} is scaled by $10^2$.}
\begin{center}
\begin{threeparttable}
\begin{tabular}{ccccccccccccc}
\toprule
\multirow{4}{*}{Pattern} & \multirow{4}{*}{A} & \multirow{4}{*}{ACL} & \multicolumn{5}{c}{In-distribution (\gls{corpd})} & \multicolumn{5}{c}{Out-of-distribution (\gls{corpdfs})}\\
\cmidrule(lr){4-8}
\cmidrule(lr){9-13}
&   &   & \multirow{2}{*}{ZF} & \multirow{2}{*}{TV} & \multirow{2}{*}{VN} & \multicolumn{2}{c}{Ours} & \multirow{2}{*}{ZF} & \multirow{2}{*}{TV} & \multirow{2}{*}{VN} & \multicolumn{2}{c}{Ours}\\
\cmidrule(lr){7-8}
\cmidrule(lr){12-13}
&   &   &   &   &   & MAP & MMSE & & & & MAP & MMSE \\
\midrule
\multirow{6}{*}{C}
& \multirow{6}{*}{4} & \multirow{3}{*}{8\%} & 27.18 & 33.07 & $\mathbf{36.92}$ & $\underline{33.22}$ & 33.15 & 26.24 & $\textbf{31.42}$ & 30.05 & 31.30 & $\underline{31.33}$ \\
&   &  & 0.73 & 0.83 & $\mathbf{0.91}$ & $\underline{0.86}$ & $\underline{0.86}$ & 0.67 & 0.73 & $\mathbf{0.76}$ & $\underline{0.74}$ & $\underline{0.74}$ \\
&   &   & 2.23 & $\underline{0.56}$ & $\mathbf{0.24}$ & 0.61 & 0.62 & 5.17 & $\mathbf{1.56}$ & 2.38 & 1.63 & $\underline{1.61}$  \\
\cmidrule(l){3-13}
&   & \multirow{3}{*}{8\%\tnote{1}} & 31.12 & 33.65 & 24.72 & $\mathbf{35.38}$ & $\underline{35.46}$ & 26.52 & $\underline{32.00}$ & 28.65 & $\mathbf{32.04}$ & 31.91 \\
&   &  & 0.81 & 0.83 & 0.67 & $\underline{0.89}$ & $\mathbf{0.90}$ & 0.70 & 0.74 & 0.70 & $\mathbf{0.77}$ & $\underline{0.76}$ \\
&   &  & 0.93 & $\underline{0.50}$ & 4.01 & $\mathbf{0.33}$ & $\mathbf{0.33}$ & 5.09 & $\mathbf{1.40}$ & 2.91 & $\mathbf{1.40}$ & $\underline{1.43}$ \\
\midrule
\multirow{3}{*}{R}
& \multirow{3}{*}{11} & \multirow{3}{*}{-} & 28.76 & 33.21 & 20.55 & $\underline{33.63}$ & $\mathbf{33.65}$ & 25.11 & $\mathbf{31.50}$ & 26.38 & $\underline{31.22}$ & 31.16 \\
&  &  & 0.74 & $\underline{0.82}$ & 0.69 & $\mathbf{0.85}$ & $\mathbf{0.85}$ & 0.61 & $\underline{0.72}$ & 0.68 & $\mathbf{0.73}$ & $\underline{0.72}$\\
&  &  & 1.56 & $\underline{0.55}$ & 10.12 & $\mathbf{0.50}$ & $\mathbf{0.50}$ & 7.13 & $\mathbf{1.54}$ & 4.90 & $\underline{1.65}$ & 1.66 \\
\midrule
\multirow{3}{*}{G}
& \multirow{3}{*}{8} & \multirow{3}{*}{-} & 32.14 & $\mathbf{34.26}$ & 23.52 & $\underline{34.07}$ & 34.00 & 26.66 & $\mathbf{31.99}$ & 28.20 & $\underline{31.74}$ & 31.58 \\
& & & 0.83 & $\underline{0.85}$ & 0.72 & $\mathbf{0.87}$ & $\mathbf{0.87}$ & 0.69 & $\underline{0.74}$ & 0.71 & $\mathbf{0.75}$ & $\underline{0.74}$ \\
&   &   & 0.73 & $\mathbf{0.43}$ & 5.20 & $\underline{0.45}$ & 0.46 & 5.40 & $\mathbf{1.38}$ & 3.33 & $\underline{1.48}$ & 1.52 \\
\midrule
\multicolumn{3}{c}{Learnable params.} & - & - & 3$\times10^{7}$ & \multicolumn{2}{c}{1578} & - & - & 3$\times10^{7}$ & \multicolumn{2}{c}{1642} \\
\bottomrule
\end{tabular}
\begin{tablenotes}
\item[1] horizontal;
C: Cartesian, R: Radial, G: 2D Gaussian, A: Acceleration 
\end{tablenotes}
\end{threeparttable}
\end{center}
\label{tab:pi_corpd_fs}
\end{table}

\begin{figure*}[!ht]
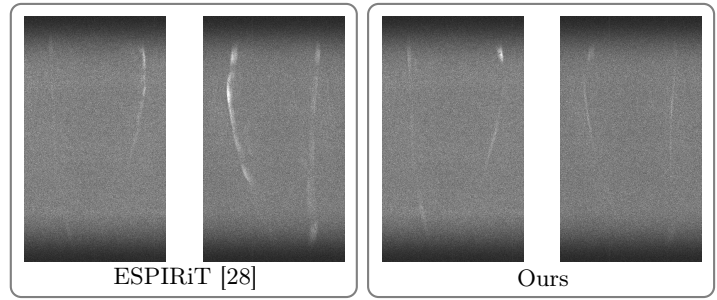

    \centering
    \def\wwidth{3.75cm}
    \begin{tikzpicture}
        \foreach[count=\isens] \senspath/\sname in {
        espirit/ESPIRiT~\cite{uecker2014espirit},
        ours/Ours
        }{
            \pgfmathsetlengthmacro{\yy}{\vpad)}
            \pgfmathsetlengthmacro{\xx}{(\isens * 2 - 1)* (\wwidth / 2 + 5.0mm) -0.75cm}
            
		  \node [inner sep=0, outer sep=0] at (\xx, \yy)  {\includegraphics[angle=0,origin=c,width=\wwidth / 2]{figures/null_space_residuals/\senspath_residual_29.png}};
          
            \pgfmathsetlengthmacro{\xx}{(\isens * 2)* (\wwidth / 2 + 5.0mm) - 0.75cm}
            \node [inner sep=0, outer sep=0] at (\xx, \yy)  
            {\includegraphics[angle=0,origin=c,width=\wwidth / 2]{figures/null_space_residuals/\senspath_residual_14.png}};
            \node at (\xx - \wwidth / 2 + 7.0mm, -\yy - 1.85 cm) {\sname};
        }
    \draw [gray, thick, rounded corners] (0.5cm, -2.10cm) rectangle (5.1cm, 1.80cm);
    \draw [gray, thick, rounded corners] (5.25cm, -2.10cm) rectangle (9.85cm, 1.80cm);
    \end{tikzpicture}
    \caption{\gls{rss} nullspace residuals for Cartesian subsampling. In each block, the measured data has 8\% \glspl{acl} (left) and 4\% \glspl{acl} (right) respectively.}
    \label{fig:rss_nullspace}
\end{figure*}

Finally, we compare the estimated sensitivity maps qualitatively to ESPIRiT \cite{uecker2014espirit} by showing the \gls{rss} null-space residuals in Figure \ref{fig:rss_nullspace}, where any residual signal points to a suboptimal estimation.

\noindent
We observe that \cite{uecker2014espirit} produces slightly better estimates than our method when more \glspl{acl} are available but falls off as the calibration region gets smaller.
In contrast, our method still gives a robust estimate of the sensitivities in all cases.

\section{Conclusion}
\label{sec:conclusion}
In this paper, we pair an analytically tractable product of Gaussian mixture diffusion model with classic smoothness penalties to tackle the non-linear \gls{mri} reconstruction problem in the diffusion framework.
The proposed method delivers satisfactory results on single- and multi-coil data, allows for fast inference, and demonstrates robustness to changes in the forward model and contrast.
A drawback of the algorithm is that it requires the tuning of many hyperparameters.
An interesting direction for future work includes developing a diffusion prior for $p_{X,\Sigma}$, not assuming a factorization, consequently aligning the method better with recent algorithms for posterior sampling with diffusion models.

\bibliographystyle{splncs04}
\bibliography{references}
\end{document}